\documentclass[journal,twoside,web]{ieeecolor}
\pdfoutput=1
\usepackage{tmi}
\usepackage{cite}
\usepackage{amsmath,amssymb,amsfonts}
\usepackage{algorithmic}
\usepackage{graphicx}
\usepackage{textcomp}
\usepackage{tabularx}  
\usepackage{booktabs}  
\usepackage{multirow}   
\usepackage{url}
\usepackage{epstopdf}
\makeatletter
\let\NAT@parse\undefined
\makeatother
\usepackage[colorlinks,
            linkcolor=blue,
            anchorcolor=blue,
            citecolor=blue]{hyperref}
\usepackage{cleveref}

\begin{document}
\title{A multi-stage semi-supervised learning for ankle fracture classification on CT images}
\author{Hongzhi Liu, Guicheng Li, Jiacheng Nie, Hui Tang, Chunfeng Yang, Qianjin Feng, Hailin Xu, Yang Chen, ~\IEEEmembership{Senior~Member, IEEE}
\thanks{This work was supported in part by the National Key Research and Development Program of China (No. 2021ZD0113202 and No. 2022YFC2504302), in part by the State Key Project of Research and Development Plan under Grant 2022YFC2401600, in part by the National Natural Science Foundation of China under Grant T2225025, in part by the Key Research and development Programs in Jiangsu Province of China under Grant BE2021703 and BE2022768, in part by China Scholarship Council under Grant 202106090126. Hongzhi Liu and Guicheng Li contribute equally in this paper. Qianjin Feng and Hailin Xu are the co-corresponding author and Yang Chen is the corresponding author.}
\thanks{Hongzhi Liu, Jiacheng Nie, Hui Tang and Chunfeng Yang are with the School of Computer Science and Engineering, Southeast University, Nanjing 210096, China.}
\thanks{Guicheng Li and Hailin Xu are with the Peking University People's Hospital, Beijing 100000, China (e-mail: xiaxi@pku.edu.cn).}
\thanks{Qianjin Feng is with the Guangdong Provincial Key Laboratory of Medical Image Processing, School of Biomedical Engineering, Southern Medical University, Guangzhou 510515, China (e-mail: fengqj99@fimmu.com).}
\thanks{Yang Chen is with the Key Laboratory of New Generation Artificial Intelligence Technology and its Interdisciplinary Applications (Southeast University), Ministry of Education, China, Jiangsu Provincial Joint International Research Laboratory of Medical Information Processing, School of Computer Science and Engineering, Southeast University, Nanjing, 210096, China, Jiangsu Key Laboratory of Molecular and Functional Imaging, Department of Radiology, Zhongda Hospital, Southeast University, Nanjing 210009, China (e-mail: chenyang.list@seu.edu.cn).}
}

\maketitle

\begin{abstract}
Because of the complicated mechanism of ankle injury, it is very difficult to diagnose ankle fracture in clinic. In order to simplify the process of fracture diagnosis, an automatic diagnosis model of ankle fracture was proposed. Firstly, a tibia-fibula segmentation network is proposed for the joint tibiofibular region of the ankle joint, and the corresponding segmentation dataset is established on the basis of fracture data. Secondly, the image registration method is used to register the bone segmentation mask with the normal bone mask. Finally, a semi-supervised classifier is constructed to make full use of a large number of unlabeled data to classify ankle fractures. Experiments show that the proposed method can segment fractures with fracture lines accurately and has better performance than the general method. At the same time, this method is superior to classification network in several indexes.
\end{abstract}


\begin{IEEEkeywords}
Ankle fracture diagnosis, Deep learning, Image classification.
\end{IEEEkeywords}

\section{Introduction}
\label{sec:introduction}
\IEEEPARstart{T}{he} ankle joint is vulnerable to injury in daily life, accounting for approximately 3.9\% of total body fractures. The incidence of ankle fractures is 187 per 100 000, 25\% of which require surgical treatment~\cite{bielska2019health}. Due to the structural complexity of ankle joint and the diversity of ankle fracture types, the treatment for ankle fractures is difficult. Recent studies have shown that the error rate in the diagnosis and treatment of ankle fractures ranks first in orthopedics trauma~\cite{walsh2018ankle}. Inappropriate treatments may cause ankle joint instability and mismatch of articular surfaces, followed by joint pain, limited mobility, early traumatic degeneration, and other problems.


X-ray imaging is still playing a key role in the diagnosis of fractures due to its low price, fast examination speed, and low radiation~\cite{costelloe2013radiography}. However, due to the extremely complex process required for imaging diagnosis and the wide range of knowledge that doctors need to learn, machine learning has broad application prospects in the field of medical imaging diagnosis. Combining machine learning with clinical diagnosis in orthopedic department can effectively improve the effect of digital medical treatment in orthopedic.

\begin{figure}[!t]
\centerline{\includegraphics[width=\columnwidth]{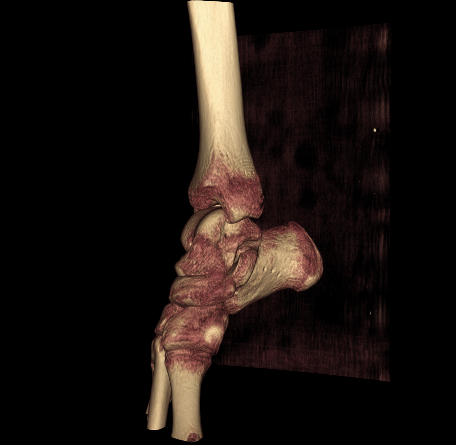}}
\caption{An example of a local injury in an ankle fracture patient.}
\label{fig1}
\end{figure}

The application of machine learning in orthopedic imaging diagnosis has been extensively studied by many researchers in recent years. In the field of orthopedic disease diagnosis, Yu~\emph{et al.}~\cite{yu2015lumbar} used SVM~\cite{cortes1995support} to classify lumbar spine ultrasound images in 2015 and realized the function of automatic recognition of bone images and interspinous images. In 2017, Kadoury~\emph{et al.} also realized the diagnosis of spinal images by machine learning~\cite{kadoury20173}. Dam~\emph{et al.} have identified targets on MRI images of knee injuries and used machine learning algorithms for classification~\cite{dam2015automatic}. Although the classification results were relatively ideal, the patients must take MRI. MRI has the characteristics of high radiation, timeconsuming, and high cost. Chen et al.10used U-NET to diagnose osteoporosis in X-ray images and performed a good result, which proved the feasibility of fracture diagnosis based on X-ray images in machine learning.

At present, there is little research on machine learning in the diagnosis of ankle fractures. Kitamura~\emph{et al.}~\cite{kitamura2019ankle} used convolutional neural networks and Pinto~\emph{et al.}~\cite{pinto2019structured} used structure report data to complete the X-ray ankle fracture recognition in 2019. The current research on the intelligent diagnosis of ankle fracture is still limited to the identification of the presence of fracture, without the identification of fracture region and type.

To solve the above problems, a novel ankle fracture classification algorithm was proposed in this study, and a new dataset was constructed based on the image data of fracture patients collected by this method. Then, we used a deep learning algorithm to segment the tibia and fibula, and registered the segmentation mask containing the fracture with the normal bone mask to preprocess and classify the X-ray image of the ankle fracture. The pre-processing process based on bone segmentation and registration can effectively segment bone contours, and make full use of unlabeled data and adopt semi-supervised network training strategy to improve the accuracy of fracture classification. Finally, the feasibility of the proposed algorithm in the field of intelligent diagnosis of ankle fracture is verified by the experimental results.

\section{Related works}
\label{sec:rw}

\subsection{Ankle fracture classification}

Some methods can be used to define ankle fracture. The first classification methodology for ankle fractures was developed by Percival Pott apud Budny and Young~\cite{budny2008analysis}, which described the number of fractured malleoli, stratifying the lesions as unimalleolar, bimalleolar, or trimalleolar. Although this classification is intuitive and easy to reproduce, it does not distinguish stable and unstable injuries, nor does it guide treatment~\cite{alexandropoulos2010ankle}. Lauge-Hansen~\cite{lauge1949ligamentous}, through cadaveric experiments, proposed a classification system that correlates the lines of ankle fractures with certain trauma mechanisms. The fractures are classified into four groups: supination-adduction, supinationeversion, pronation-eversion, and pronation-abduction. The first term indicates the position of the foot at the time of injury and the second refers to the direction of the force applied to the foot at the time of the trauma~\cite{lauge1949ligamentous,pimenta1991estudo,russo2013ankle}. According to this classification, the supine-eversion pattern is the most frequent in emergency departments, with a prevalence ranging from 40\% to 75\%~\cite{hamilton2012traumatic}.

Danis~\cite{danis1949fractures} and Weber~\cite{weller1967weber} proposed another classification system based on the localization of the main fibular fracture line, dividing the fractures into three groups: type A (below the syndesmosis level), type B (at the syndesmosis level), and type C (above the syndesmosis). Despite its simplicity and ease-ofreproduction, this classification does not consistently predict the extent of the injury in the tibiofibular syndesmosis, as several studies have already demonstrated, since type B and C fractures can be treated in a similar way regardless of the location, according to the presence or absence of ligament instability at the site. This classification also disregards the state of the structures on the medial side, a vital osteoligament structure, and it is not possible to compare prognosis, treatment, or evolution of the pathology with this classification alone~\cite{broos1991operative,nielson2005radiographic}. The AO-OTA group expanded the Danis-Weber classification scheme, developing a classification based on the location of the fracture lines and on the degree of comminution. Thus, it allows describing the severity and degree of instability associated with a specific fracture pattern~\cite{meinberg2018fracture,Mller1990TheCC}. 


In 2020, Olczak and colleagues successfully applied a CNN for ankle fracture classification~\cite{olczak2020ankle} using the image-level classification model ResNet~\cite{he2016deep}, but without automated delineation of the fracture. Ideally, CNNs should combine object detection with segmentation, and thus offer localization and classification simultaneously—for example to better guide junior doctors during their early learning curves by presenting an exact visual outline of the fracture line itself. In addition, CNNs are often trained with large datasets without selecting cases that facilitate the most efficient training rate for the CNN. This results in a large portion of unnecessarily labelled and/or annotated cases, because these contribute minimally to the performance of the model.

\begin{figure*}[htbp]
\centering
\centerline{\includegraphics[width=\textwidth]{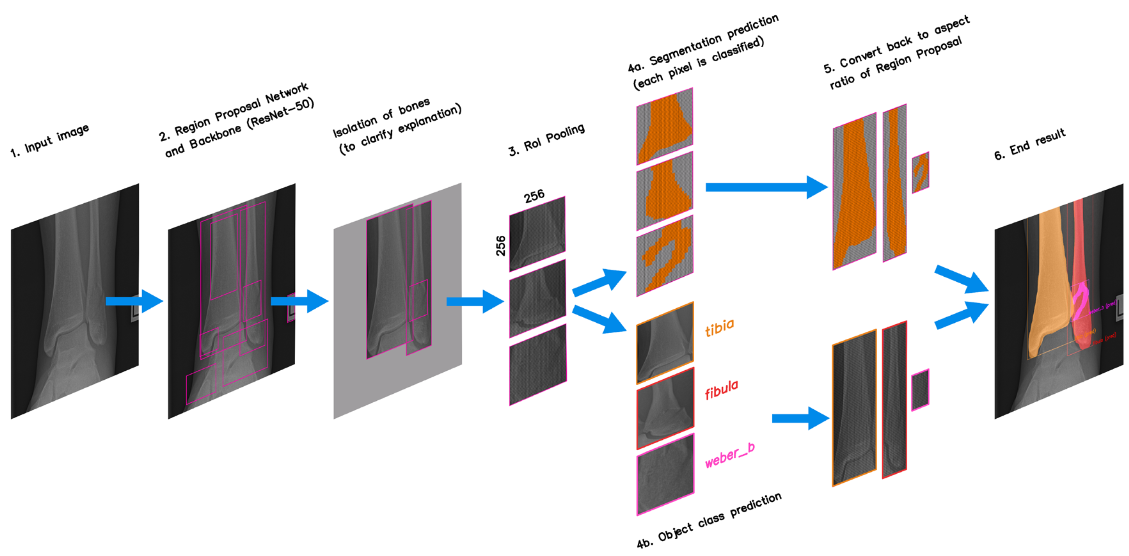}}
\caption{The flowchat shows the process from input raw CT images to segmentation masks. }
\label{fig2}
\end{figure*}

\subsection{Medical image segmentation}

Recently, due to the success of the deep convolution network in image processing and it has achieved widespread attention in medical image segmentation~\cite{zhang20222k,wang2022eanet}. U-Net~\cite{ronneberger2015u} consists of an encoder path and a decoder path, which gives it the u-shaped architecture. U-Net++\cite{zhou2018unet++} is essentially a deeply supervised encoder-decoder network where the encoder and decoder sub-networks are connected through a series of nested, dense skip pathways. Receptive fields of U-Nets may be enlarged by adding more downsampling and convolutional layers. However, this increases the number of parameters and adds the risk of overfitting. Existing approaches propose to directly enlarge the receptive fields based on CNN features to capture features with long-range dependency. Following the design principle of large-size kernels, Peng~\emph{et al.}~\cite{peng2017large} designed large kernels to capture rich global context information. Chen~\emph{et al.}~\cite{chen2017deeplab} effectively enlarged the field of view of filters to incorporate a larger context without increasing the number of parameters or the amount of computation. Zhao~\emph{et al.}~\cite{zhao2017pyramid} exploited the capability of global context information by different-region-based context aggregation through a pyramid pooling module.

\subsection{Semi-supervised classification}

Given a training dataset that consists of both labeled instances and unlabeled instances, semi-supervised classification aims to train a classifier from both the labeled and unlabeled data, such that it is better than the supervised classifier trained only on the labeled data. Reducing the number of manual annotations required to train medical imaging models will significantly reduce both the cost and time required for model development, making automated systems more accessible to different specialties and hospitals, thereby reducing workload for radiologists and potentially improving patient care. The pseudo-labeling methods differ from the consistency regularization methods in that the consistency regularization methods usually rely on consistency constraint of rich data transformations. Consistency regularization is based on the manifold assumption or the smoothness assumption, and describes a category of methods that the realistic perturbations of the data points should not change the output of the model~\cite{oliver2018realistic}. Pseudo-label~\cite{lee2013pseudo} proposes a simple and efficient formulation of training neural networks in a semi-supervised fashion, in which the network is trained in a supervised way with labeled and unlabeled data simultaneously. Hybrid methods combine ideas from the above-mentioned methods such as pseudo-label, consistency regularization and entropy minimization for performance improvement. MixMatch~\cite{berthelot2019mixmatch} combines consistency regularization and entropy minimization in a unified loss function. This model operates by producing pseudo-labels for each unlabeled instance and then training the original labeled data with the pseudo-labels for the unlabeled data using fully-supervised techniques.

\section{Method}
\label{sec:md}

\subsection{Overview}

The key observation on which our approach is based is that the classification of ankle fractures is related to the relative position of the fracture line and the joint tibiofibular region, and that there are differences in position when different fractures occur. Based on this idea, we divided the classification of ankle tibiofibular fractures into two main steps.

In the first stage, an convolutional neural network is used to identify the tibia and fibula in the source CT scan images and segment it into two anatomical structures: tibia and fibula bone.   Then the symmetric structure of the ankle is registered to obtain a spatial transformation matrix via rigid registration algorithm, which is used to compute the transformed masks from the healthy ankle masks. Then the same region was cropped on the registered masks based on the predetermined tibia-fibula joint area on the health mask.

In the second stage, A semi-supervised network was used to learn ankle fracture classification from labeled and unlabeled data. A pre-trained backbone was used to extract features and generate pseudo-labels for unlabelled data. Then the relational weight network based on the squeeze-and-excitation network was trained to calculate the non-linear distance metrices between labelled and unlabelled samples, in order to improve the accuracy of pseudo-labelling.

\subsection{Tibia and fibula segmentation network}

\begin{figure}[htbp]
\centering
\centerline{\includegraphics[width=0.5\textwidth]{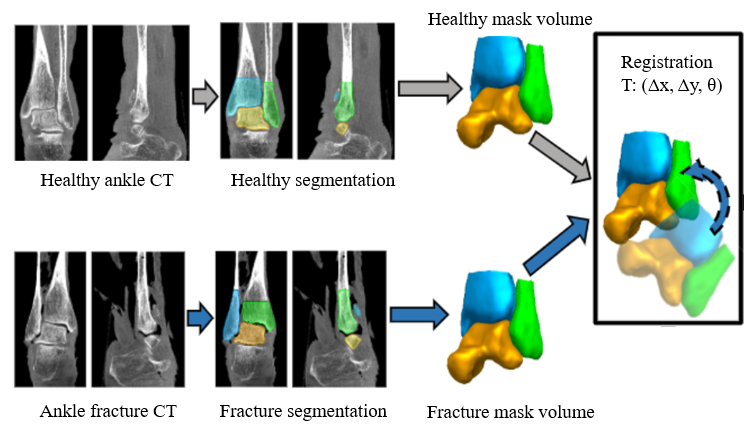}}
\caption{Healthy and fracture ankle joint segmentation mask registration.}
\label{fig3}
\end{figure}

The tibia and fibula segmentation model in this study can segment a single tibia and fibula object by combining object bounding box and semantic segmentation (Fig.~\ref{fig2}). A CT image containing multiple slices (Fig.~\ref{fig2}.1) is used as input to the network. ResNet-50 acts as a backbone to create a number of bounding boxes, each of which is a specific tibia or fibula classification object (Figure Fig.~\ref{fig2}.2), in conjunction with the Region Proposal Network (RPN). The RPN and backbone create countless bounding boxes, where each box has a high likelihood of presence of tibia and fibula. The boundary boxes of the same category were merged so that the tibia and fibula could be separated from the CT images. In this way, fracture fragments belonging to the tibia or fibula can be found in the image. The algorithm overhead of the subsequent segmentation process can be reduced by clipping the bone region. Adjust the size of each region proposal through the Region of Interest (RoI) pooling operation to fit a fixed height and width size of $256 \ × $256 (Fig.~\ref{fig2}.3). These cropped images are then used to simultaneously segment and classify. The network performs feature learning for each pixel in the region suggestion to create a segmentation mask (Fig.~\ref{fig2}.4a) , while each region suggestion classifies the segmentation object (Fig.~\ref{fig2}.4b). Finally (Fig.~\ref{fig2}.5) the predicted results are then resized to the original height and width dimensions and projected onto the output image.



\subsection{Ankle masks registration}

After tibia-fibula segmentation is completed, the next stage is to register the masks of the injured ankle with the normal masks then syndesmosi regions are cropped for fracture type determination. This will be done by using the healthy ankle masks as the guideline for transforming fractured ankle masks. The image pairs of potential areas on the injured ankle masks and healthy ankle masks are extracted in the image domain. The former ones are used as the source point cloud and the latter ones are used as the target point cloud as shown in Fig.~\ref{fig3}. Healthy and fracture segmentation results were obtained in the tibia-fibula segmentation network from the health ankle CT and ankle fracture CT, respectively. The healthy and fracture mask volumes will be mirrored and flipped, and then the rigid registration from the injured to the healthy ankle will be computed. Next, the Iterative Closest Point (ICP)~\cite{besl1992method} algorithm is applied to compute the spatial transformation matrix $\mathcal{T}$, which is used to rotate and translate the flipped source image to obtain the transformed image that is aligned with the source image in the syndesmosi region. The mask pairs of the tibiofibular syndesmosis are extracted by cropping both the source and transformed masks according to the bounding box coordinates of the potential area. 


\subsection{Semi-supervised fracture classification}

\begin{figure*}[htbp]
\centering
\centerline{\includegraphics[width=\textwidth]{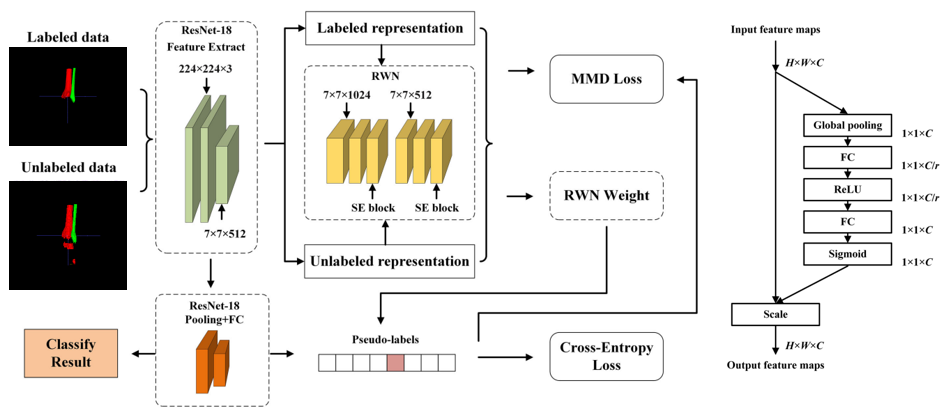}}
\caption{The semi-supervised model for ankle mask classification based on the self-training strategy.}
\label{fig4}
\end{figure*}

A semi-supervised convolutional neural network learns to determine the type of ankle fracture using limited labeled data and some unlabeled data is shown in Fig.~\ref{fig4}. The ResNet~\cite{he2016deep} is used to extract features from image samples. The accuracy of feature representation can be further improved by the Squeeze-and-Excitation network (SENet)~\cite{hu2018squeeze} network structure in a relational weight network (RWN), and the feature relationships between ankle fracture masks can be effectively learned. Coalition training these two networks at the same time can avoid the erroneous iterations of single network training. Therefore, the RWN is to compute the non-linear distance metrics of the samples based on the pre-trained feature extraction network ResNet-18, and then calculate the relational weights of unlabelled samples and prototypes of classes generated from labelled samples using RWN to reduce the noise of pseudo-labels generated by the feature extraction network. The weighted pseudo-label samples are used to calculate the loss function. Finally, the maximum mean difference (MMD) loss is added to the loss function to reduce the statistical distribution distance between the unlabelled set and the labelled set.

The SENet is used to calculate the interdependence between convolution channels. The characteristic response of the channel is adaptively recalibrated, to re-fuse the information characteristic constructed by the channel during the convolution process. The network structure of the SE block is shown on the right in Fig.~\ref{fig4}. To obtain the global receptive fields of the channel, the SE block first uses global average pooling to represent the channel, then automatically estimate the weight of each channel through squeezing and excitation to focus on key channels and suppress irrelevant channels for different tasks. In this paper, we used the relation module~\cite{zhang2018relationnet2} to calculate the non-linear distance metric between labelled data and unlabelled data to solve the semi-supervised classification problem. For feature extraction, we used the ResNet-18, which is pre-trained on the ImageNet dataset~\cite{deng2009imagenet}. RWN consists of four convolutional layers and two SENet blocks. The feature representations of the unlabelled masks and labelled masks obtained by ResNet-18 are spliced and input into RWN. The feature representations of unlabelled and labelled samples obtained by ResNet blocks are spliced and then put into RWN. Only in the first convolutional layer downsampling is used with the stride of 2, and BatchNorm is performed after each convolution operation. ReLU activation function is used in the first and third convolutional layers. SENet blocks are used to calculate the correlation between channels more accurately. Finally, the average pooling layer and the fully connected layer are used to generate relationship scores.

After building the network, we first load the pre-trained ResNet-18 parameters, and then randomly initialize the remaining parameters. The labelled samples and unlabelled samples are fed into the ResNet-18 network. From the last Softmax layer, the predicted labels of the labelled sample and the pseudo labels of the unlabelled samples are obtained. We then calculate the prototypes of classes of each class according to the true labels of the labelled samples, and concatenate the feature vectors of labelled samples with each class prototypes. We input the spliced feature vector into RWN and then calculate the RWN loss based on the relationship weight obtained by RWN and the real label. The RWN network parameters are optimized by minimizing the RWN loss function. We then calculate the prototypes of classes of each class according to the true labels of the labelled samples, and concatenate the feature vectors of labelled samples with each class prototypes. We input the spliced feature vector into RWN and then calculate the RWN loss based on the relationship weight obtained by RWN and the real label. The RWN network parameters are optimized by minimizing the RWN loss function. Then we concatenate unlabelled samples with prototypes of classes. RWN is used to calculate the relation weight between the unlabelled samples and the prototypes of classes, and the pseudo-label obtained by ResNet-18 is weighted according to the calculated weight. The weighted pseudo-label is used to calculate the Softmax loss. The feature representation obtained by ResNet-18 is selected according to the weight, and the selected feature representation of the unlabelled sample is used to calculate the MMD loss. ResNet-18 network parameters are then optimized by minimizing the total loss function. When evaluating the trained model, the test sample is used as input, and the retrained ResNet-18 is used to directly output the prediction classification. We finally compare the predicted result with the true label of the sample to calculate the accuracy of the test.

We calculate the loss function for labelled ankle masks and unlabelled masks respectively. For labelled samples, the cross-entropy loss is calculated directly to obtain the supervised loss $\mathcal{L}_l$. For unlabelled samples, pseudo-labels weighted by $w_i$ are used to calculate unsupervised losses $\mathcal{L}_u$. The loss functions $\mathcal{L}_l$ and $\mathcal{L}_u$ are defined as:

\begin{align}
    \mathcal{L}_l(x_l;[\theta_{R}, \theta_s, \phi]) = \mathcal{L}_{ce}(f_{\theta_s,\phi}(x_i),y_i)  (x_i,y_i)\in x_l, \label{loss1} \\
    \mathcal{L}_u(x_u;[\theta_{R}, \theta_s, \phi]) = \mathcal{L}_{ce}(w \odot f_{\theta_s,\phi}(x_i),y_i)  (x_i,y_i)\in x_l, \label{loss2} 
\end{align}

\noindent where $f_{\theta_s,\phi}(\cdot)$ is the feature extractor and classifier parameters. The network uses the alternate training form of RWN and feature extraction network to update network parameters. $\mathcal{L}_{ce}$ is the cross-entropy loss function, which calculates loss in a standard way on $x_l$. For the pseudo-label generated by xu, its predicted value is weighted by $w_i$ before entering the Softmax layer.

And the most confident predictions were used for calculations. The resulting sets are respectively $T_l$ and $T_u$. Here, we also set up a buffer to store the recently selected confident examples. The sets expanded by the buffer are $T_l^*$ and $T_u^*$. The network parameters are then optimized by minimizing the MMD between them. The loss function LMMD is defined as:

\begin{equation}
  \begin{aligned}
    \mathcal{L}_{MMD} & = MMD(T_l^*, T_u^*)  \\
    & = \parallel \frac{1}{m}\sum_{i=1}^m k(x_l^i) - \frac{1}{n}\sum_{j=1}^m k(x_u^j)    \parallel^2
  \end{aligned}\label{loss3}
\end{equation}

\noindent where $x_i^l$ is the feature representation of the sample in $T_l^*$, $x_u^j$ is the feature representation of the sample in $T_u^*$, and $k(\odot)$ is the Gaussian radial basis function. The feature extraction network uses a combination of losses to optimize network parameters. Therefore, the expression of the total loss function $L_{total}$ with the balance parameter $\lambda$ of the feature extraction network is defined as:

\begin{align}
    \mathcal{L}_{total} & = \mathcal{L}_{l} + \mathcal{L}_{u} + \lambda \mathcal{L}_{MMD}. \label{loss4}
\end{align}

Thus this semi-supervised fracture classification network use a self-training technique based on Squeeze-and-Excitation network (SENet) to learn the non-linear distance metrics between labeled and unlabeled mask data.

\section{Experimental results}
\label{sec:er}

\subsection{Datasets and preprocessing}

Our proposed approach was validated in the two datasets to evaluate the effectiveness that are tibia-fibula segmentation dataset and ankle fracture classification dataset. They were obtained from Peking University People's Hospital\footnote{https://english.pkuph.cn/index.html} in this study.

The tibia-fibula segmentation dataset consists of 176 CT scans of clinical cases acquired by Philips MX 16-slice spiral CT (X-ray tube current is 415 mA and kilovolt peak is 140 kV) and each scan includes annotations of tibia and fibula. The CT radiographs size varies from 512$\times$512$\times$368 to 768$\times$768$\times$413, with a average value of 679$\times$679$\times$394. The spatial resolution of CT scanning varies from 0.24$\times$0.24$\times$0.40 to 0.45$\times$0.45$\times$1.25 $mm^3$, with a average value of 0.36$\times$0.36$\times$0.45 $mm^3$. The ground truth segmentation were annotated by expert orthopedic surgeons and experienced radiologists via 3D Slicer~\cite{kikinis20133d} and ITK-SNAP~\cite{yushkevich2006user}. Bones separated by fracture on the tibia or fibula are considered to be components of the original skeleton and need to be correctly segmented. We follow a similar splitting mode of the Medical Segmentation Decathlon dataset~\cite{simpson2019large} and the dataset was randomly partitioned into 122 training set and 55 testing set. Then intensity of CT radiographs is normalized to the range of [0, 255]. The input image patches cropped from CT images to the segmentation network were resized to 64$\times$64$\times$64. We also did a resampling operation across all input volumes to a common voxel spacing of 1.5$\times$1.5$\times$10 $mm^3$. To enhance the variability of the data representation, a light data augmentation techniques on the data and ground truth labels were applied. Rotation, random translation, horizontal and vertical flipping were used for data augmentation. On a random basis, the data was rotated between $-15$ to $+15$ degree, and scaled between $0.9-1.1$ range. This ensured slight robustness and variability in training the network.

The ankle fracture classification dataset contains a total of 612 ankle radiographic CT from different patient studies, including 330 labeled data and 282 unlabeled data. The size of CT images varies from around 512$\times$512$\times$301 to 768$\times$768$\times$413 and the spatial resolution from 0.21$\times$0.21$\times$0.40 to 0.51$\times$0.51$\times$1.25 $mm^3$. We resize all the images to a uniform size of 512$\times$512 pixels for network training and validation. The fractures of interest in this dataset are the lateral malleolar fractures corresponding to the level of the fracture in relation to the ankle joint, measured as type A, B, or C by the Danis-Weber and AO-OTA fracture classification system~\cite{danis1949theorie,weller1967weber,orthopaedic1996fracture,sanders2018ota}. Based on the localization of the main fibular fracture line, researchers defined type A as below the syndesmosis, type B as being at the level of syndesmosis and type C as above the syndesmosis. Three experienced orthopedic surgeons were asked to annotate and review the fracture area in each case then provided ground truth fracture classification labels for 330 cases of data. 285 cases of these data were split as the training set and 45 cases for testing, where the three types of fracture data were nearly balanced in testing set. The rest 282 cases of data without ground truth labels were used for semi-supervised learning and different proportions of unlabelled data were used during the model training process. Therefore, it can save time in manual labelling of fracture data and the classification method that makes full use of labeled data and unlabeled data is needed.

\subsection{Implementation details}

The proposed method was implemented using the PyTorch framework, and the training and testing phases of all experiments were executed on a workstation with an NVIDIA GeForce RTX 3090 GPU. The proposed framework uses CT images as input to obtain the ankle fracture classification after three stages, which are tibia-fibula segmentation, segmented masks registration and semi-supervised classification. The tibia-fibula segmentation network was trained with stochastic gradient descent (SGD) optimization with Nesterov momentum of 0.99. The initial learning rate was 1e-4 and the batch size was set to be 2 for 500 epochs. To promote the convergence of the networks, the learning rates were periodically reduced by $0.1$. 

The healthy ankle masks with no fracture and segmented masks of fracture bones from the tibia-fibula segmentation network are the input pairs for the rigid registration algorithm. The healthy unfractured masks were used as the target point cloud, and the segmented fractured masks were used as the source point cloud. An affine transform consisting of translation, rotation, and scaling operations was computed using the iterative closest point (ICP) algorithm. The masks within the syndesmosis region of the registered masks were used for fracture classification. The maximum number of iterations was 50 and the transformation epsilon of convergence threshold was set to be 1e-8.

The semi-supervised classification network makes full use of limited labeled data and many unlabeled data for determining fracture types. It also use SGD with momentum as the optimizer to alternately train the target model for 500 epochs. The momentum rate is set to 0.9. The initial learning rate is 1e-3, and the batch size of the labelled and unlabelled data sets is 32. In the consistency training, the sample selection threshold is set to 0.5, and the balance parameter $\lambda$ of the loss function is set to 15.

In training phase, rotation, shifting, random crops, random horizontal and vertical flipping, as well as voxel intensity jittering were applied as on-the-fly data augmentation. Model training and hyperparameter tuning were performed only on the training set. The voxel intensity of all scans was truncated within the Hounsfield Unit (HU) window of $[-1000, 1000]$ and normalized to $[0, 1]$. The model that achieved the best validation results was chosen and tested on the testing set for objectivity.

\subsection{Evaluation metrics}

For quantitative evaluation in tibia-fibula segmentation, we measure the Dice Similarity Coefficient (DSC) and the 95-th percentile of Hausdorff Distance (HD95) between segmentation results and the ground truth in 3D space. Dice falls in [0, 1], and a larger value indicates a better segmentation.


\begin{align}
    Dice(\mathcal{R}_a, \mathcal{R}_b) &= \frac{2|\mathcal{R}_a \cap \mathcal{R}_a|}{|\mathcal{R}_a|+|\mathcal{R}_b|}, \label{m1} \\ 
    HD'(\mathcal{S}_a, \mathcal{S}_b) &= \max_{i\in \mathcal{S}_a}\min_{j\in \mathcal{S}_b}\parallel i-j \parallel_2,\\
    HD(\mathcal{S}_a, \mathcal{S}_b) &= \max\{ HD'(\mathcal{S}_a, \mathcal{S}_b), HD'(\mathcal{S}_b, \mathcal{S}_a)\}, \label{m2}
\end{align}

\noindent where $\mathcal{R}_a$ and $\mathcal{R}_b$ in Eq.~\ref{m1} are the segmentation mask for a certain class and the corresponding ground truth label, respectively. and $\mathcal{S}_a$ and $\mathcal{S}_b$ represent the set of surface points of one organ segmented by a CNN and the ground truth respectively. HD95 is similar to HD, and it uses the 95-th percentile instead of the maximal value in Eq.~\ref{m2}.


For a comprehensive evaluation of fracture classification, we computed the Accuracy (Acc), Precision (Pre), Specificity (Spe), Sensitivity (Sen), and area under curve (AUC). These evaluation metrics are defined as follows: 

\begin{align}
    Accuracy &= \frac{N_{TP}+N_{TN}}{N_{TP}+N_{FP}+N_{TN}+N_{FN}},\\
    Precision &= \frac{N_{TP}}{N_{TP}+N_{FP}},\\
    Specificity &= \frac{N_{TN}}{N_{TN}+N_{FP}},\\
    Sensitivity &= Recall = \frac{N_{TP}}{N_{TP}+N_{FN}},\\
    AUROC &= \frac{\sum_{s_i\in P} rank_{s_i}-N_P \times (N_P + 1)/2}{N_P\times N_N},
\end{align}\label{m3}

\noindent where $N_{TP}$, $N_{TN}$, $N_{FP}$, $N_{FN}$ denote the numbers of true positive, true negative, false positive, and false negative samples, respectively. $N_{P}$ and $N_{N}$ denote the numbers of positive and negative samples, respectively. And $rank_{s_i}$ is the rank of the $s_i$-th positive sample.

\begin{figure*}[htbp]
\centering
\centerline{\includegraphics[width=\textwidth]{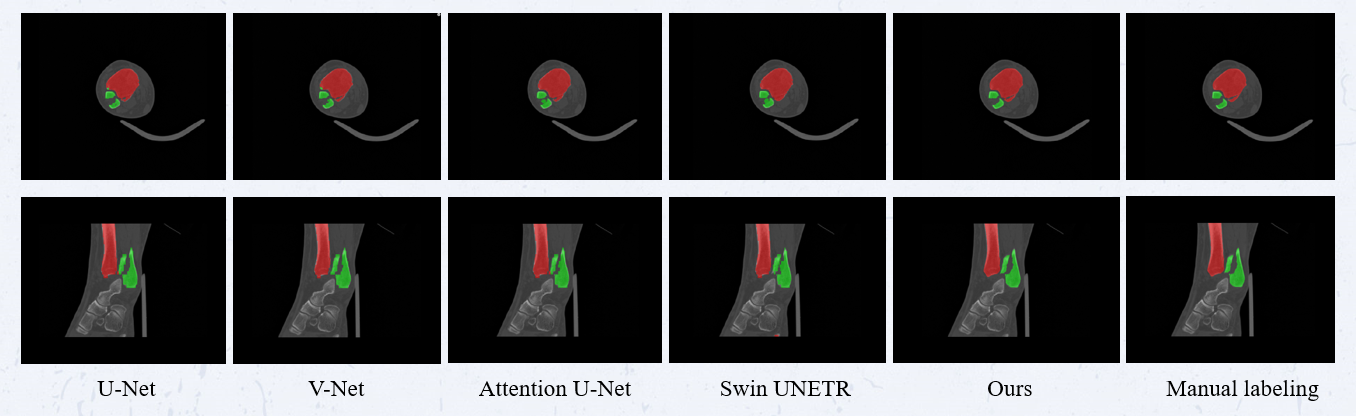}}
\caption{Comparison of qualitative results of different segmentation methods. The red is the fibula and the green means tibia. }
\label{fig5}
\end{figure*}

\begin{table*}
\caption{Quantitative results of different methods in the bone segmentation task.}
\centering
\begin{tabular}{|c|c|c|c|c|c|c|}
\hline
Method & Tibia Avg Dice & Fibula Avg Dice & Tibia Avg HD95 & Fibula Avg HD95 & Mean Dice & Mean HD95 \\
\hline
U-Net~\cite{ronneberger2015u} & 0.6101 & 0.5061 & 64.9172 & 98.1296 & 0.6081 & 86.5234\\
V-Net~\cite{milletari2016v} & 0.7106 & 0.5926 & 55.8916& 89.1579 &0.6516 & 72.5247 \\
UNETR~\cite{hatamizadeh2022unetr} & 0.8237 & 0.7892 & 15.9422 & 56.5713 &0.8064 & 36.2567 \\
SegResNet~\cite{myronenko20193d} & 0.9268 & 0.8330 & 22.9826 & 33.8644 & 0.8799 & 28.4235 \\
Swin UNETR~\cite{hatamizadeh2021swin} & 0.9336 & 0.9085 & 9.9005 & 4.3171 & 0.9235 & 12.4010 \\
Attention U-Net~\cite{oktay2022attention} & 0.9353 & 0.8990 & 7.2843 &  4.7101 &  0.9163 &  5.9972 \\
nnU-Net~\cite{isensee2021nnu} & 0.9452 & 0.9092 & 4.0308 & 3.6195 & 0.9272 & 3.8252 \\
SAM~\cite{Kirillov_2023_ICCV} & 0.8514 & 0.8715 & 9.6352 & 7.9901 & 0.8615 & 8.8127 \\
Ours & 0.9619 & 0.9233 & 3.5848 & 3.7878 & 0.9426 & 3.6863 \\
\hline
\end{tabular}
\label{tab1}
\end{table*}

\subsection{Tibia and fibula segmentation}

The segmentation results of the proposed method are shown in Table~\ref{tab1}. We first compare our method with some single task segmentation methods. There are three kind of comparison methods. One is the traditional medical image segmentation network, including U-Net~\cite{ronneberger2015u}, V-Net~\cite{milletari2016v}, SegResNet~\cite{myronenko20193d}, Attention U-Net~\cite{oktay2022attention}, and nnU-Net~\cite{isensee2021nnu}. These methods are widely used in medical image segmentation and have achieved good results. The second category is the recent medical image analysis frameworks based on Transformer, including UNETR~\cite{hatamizadeh2022unetr} and Swin UNETR~\cite{hatamizadeh2021swin}. The second category is large visual models, including SAM~\cite{Kirillov_2023_ICCV}. In traditional networks, the Dice of U-Net~\cite{ronneberger2015u} and V-Net~\cite{milletari2016v} are both greater than 60\%, and nnU-Net~\cite{isensee2021nnu} reaches 92.72\%. In the Transformer-based network, UNETR~\cite{hatamizadeh2022unetr} and Swin UNETR~\cite{hatamizadeh2021swin} have pleased performance, their Dice is greater than 90\%. Comparing the proposed method with the existing method, our method has the best Dice and HD95, which are 94.26\% and 3.69, respectively. Although the proposed method only has a slight performance lead compared with nnU-Net~\cite{isensee2021nnu} and Swin UNETR~\cite{hatamizadeh2021swin}, it should be noted that nnU-Net~\cite{isensee2021nnu} and Swin UNETR~\cite{hatamizadeh2021swin} fail to generate continuous masks, while the proposed method can also be finely segmented around the fracture suture. It indicates that our method is more outstanding for network performance improvement. In addition, the proposed method carries out a subdivision design within the attention process. And the existing Transformers usually do not carry out more detailed design inside the attention process. This design approach could be some potential future solutions. In general, the proposed method is competitive with the existing segmentation methods, and some metrics are even better than the large visual model.

The visualized segmentation results are shown in Fig.~\ref{fig5}. The segmentation results of the proposed method can clearly show tibia and fibula in the 2D slicer and 3D view. These segmentation examples also show that the landmark task can optimize the segmentation results. The gap between the fracture bones is small, and this region is easily be missegmented. The segmentation result of U-Net~\cite{ronneberger2015u}, V-Net~\cite{milletari2016v}, and SegResNet~\cite{myronenko20193d} show that the tibia in this region have missegmented depression. Benefit from the landmark annotation of the ankle in the landmark task, our method can correctly segment this region. In addition, we analyze the results of different segmentation objectives, which are shown in Fig.~\ref{fig5}. Because fracture fragment is thin and attached to bone, the boundary contrast between the two is low, so the segmentation effect of cartilage is poor compared with bone. This is also reflected in the visualized segmentation results. It can be seen that the segmented boundary of the ankle fracture is not clear, and there are spillage and perforation. However, compared with other methods, our segmentation results reduce this phenomenon. 



\begin{table*}
\caption{Quantitative results of different supervised and semi-supervised methods with registered masks or CT images in the ankle fracture classification task.}
\centering
\begin{tabular}{c|c|c|c|c|c|c|c|c|c}
\hline
\multirow{2}{*}{Type} & \multirow{2}{*}{Method} & \multicolumn{4}{c|}{Labeled mask percentage} & \multicolumn{4}{c}{Labeled CT images percentage}\\
\cline{3-10}
 & & 5\% & 20\% & 50\% & 70\% & 5\% & 20\% & 50\% & 70\% \\
\hline
\multirow{3}{*}{Consistency based} & SRC-MT~\cite{liu2020semi} & 53.06 & 53.25 & 59.72 & 65.44  & 52.09 &  52.57 &  56.08 &  63.41  \\
 & S2MTS2~\cite{liu2021self} & 55.81 & 56.37 & 60.88  & 64.13 & 52.06  & 52.41 & 57.34  & 63.19 \\
 & Ladder~\cite{rasmus2015semi} & 55.01 & 55.69 &  61.50 & 65.19  & 52.89 & 53.05 & 56.81  &  64.92 \\
\hline
\multirow{3}{*}{Hybrid algorithm} & MixMatch~\cite{berthelot2019mixmatch} & 58.97 & 59.65 & 65.28  & 69.23  & 57.02  & 57.89 & 63.73  & 66.02  \\
 & FixMatch~\cite{sohn2020fixmatch} & 62.04 & 62.60 & 67.61  & 70.49  & 60.92 & 61.34 & 63.65  & 67.93  \\
 & ReMixMatch~\cite{Berthelot2020ReMixMatchSL} & 61.15 & 61.57 & 66.56  & 69.96  & 59.76  & 60.29 & 64.02  &  68.41 \\
\hline
\multirow{4}{*}{Pseudo label} & PL~\cite{lee2013pseudo} & 60.88 & 61.28 & 68.37  & 71.50  & 58.82 & 59.06 & 62.51  &  69.08 \\
 & ACPL~\cite{liu2022acpl} & 63.20 & 63.76 & 69.21  & 73.14  & 61.39  & 61.57 & 65.19  & 70.35  \\
 & PEFAT\cite{zeng2023pefat} & 62.08 &  62.90 & 70.57  & 74.04  & 60.89  & 61.19 & 64.53  & 70.78  \\
  & Ours  &  65.01 & 66.85 & 71.20  & 75.22  & 61.57  & 62.13 & 66.30  & 71.25 \\
 \hline
\multicolumn{2}{c|}{}& Sen  &  Spe &  Pre & Acc & Sen  & Spe  & Pre & Acc \\
 \hline
 End-to-end framework & ResNet~\cite{he2016deep} with raw CT & 28.87 & 30.31  &  41.43 & 42.22 & 21.33 & 25.10 &  36.58 & 39.78 \\
\hline
\multirow{5}{*}{Supervised method} & ResNet~\cite{he2016deep} & 76.19 & 75.00  &  81.37 & 85.43 & 72.53 & 73.76 & 80.35 &  84.26\\
 & DenseNet~\cite{huang2017densely} & 77.32 & 76.43  &  82.77 & 86.29 & 75.32 & 76.11 & 81.04 & 84.67\\
 & EfficientNet~\cite{tan2019efficientnet} & 81.50 &  83.68 &  84.51 & 85.34 & 79.08 & 83.24 & 82.18 & 83.35 \\
 & Vision Transformer~\cite{dosovitskiy2020image} & 76.61 & 75.65  & 79.66  & 82.89 & 77.31 & 78.80 & 81.12 & 81.96 \\
 & Swin Transformer~\cite{liu2021swin} &  78.62 & 77.82  & 76.86  & 84.13 & 75.65 & 76.73 & 80.55 & 83.79\\
 & Ours & 82.44 & 87.83  & 85.22  & 87.59  & 81.39  & 85.23  & 89.57  &  86.11 \\
\hline
\end{tabular}
\label{tab2}
\end{table*}

\begin{figure}[!t]
\centerline{\includegraphics[width=\columnwidth]{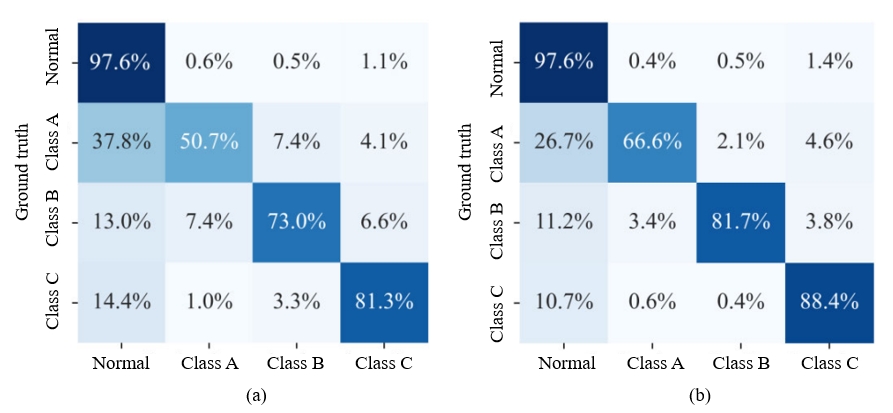}}
\caption{A comparison of the row-normalized confusion matrix between(a) a representative semi-supervised method\cite{zeng2023pefat} and (b) the proposed framework on the test set.}
\label{fig6}
\end{figure}

\subsection{Ankle fracture classification}

To test and compare the classification performance of the proposed model, we trained and tested the following semi-supervised classification methods on the ankle fracture classification dataset, including consistency based methods~\cite{liu2020semi,liu2021self,rasmus2015semi}, hybrid algorithm~\cite{berthelot2019mixmatch,sohn2020fixmatch,Berthelot2020ReMixMatchSL}, and pseudo label~\cite{lee2013pseudo,liu2022acpl,zeng2023pefat}. Among them, SRC-MT~\cite{liu2020semi}, S2MTS2~\cite{liu2021self} and Ladder~\cite{rasmus2015semi} are consistency based methods, MixMatch~\cite{berthelot2019mixmatch}, FixMatch~\cite{sohn2020fixmatch} and ReMixMatch~\cite{Berthelot2020ReMixMatchSL} are hybrid algorithm, PL~\cite{lee2013pseudo}, ACPL~\cite{liu2022acpl} and PEFAT\cite{zeng2023pefat} are pseudo label methods. In the method based on self-training and in FixMatch~\cite{sohn2020fixmatch}, the pseudo-labels used in training are generated by the model of the previous epoch. We set a hard threshold for pseudo-labels, and select pseudo-labels with confidence greater than 0.7 to participate in model training. In addition, for the fairness of comparison, all methods are trained on the pre-trained ResNet-50, and the dataset uses the same preprocessing and enhancement methods.

Table~\ref{tab2} shows the performance of these methods under three different AP. We used a fully supervised end-to-end model trained only with labelled data as the baseline. It used only a single standard ResNet-50 as feature extraction and classification, with a softmax loss function. We found that the accuracy of all methods is better than the baseline. As the number of labelled samples increases, the improvement of model performance gradually decreases, but the advantages of our proposed method gradually increases compared with other semi-supervised methods.Compared with the current more advanced semi-supervised method FixMatch~\cite{sohn2020fixmatch}, our method also has better classification performance.

To show the effectiveness of the proposed method, we compared it with the state-of-the-art algorithms. These methods have been validated on histopathological medical images of medical and are consistent with our study objectives. The methods in Table~\ref{tab2} below are supervised learning methods that only use annotated data. Comparable performance was achieved by the proposed method compared with these supervised methods. Table~\ref{tab2} also shows the accuracy of each category in the BreakHisdataset under the condition of different AP. It can be seen that as the number of labelled samples increases, the accuracy of all categories increases. Among them, when the number of labelledsamples is large (20\%), it is more difficult to classify the A and C classes than other classes; the same finding is reported by other fully supervised studies. However, when the numberof labelled samples decreases, the impact of category classifica-tion difficulty gradually decreases, and the accuracy of variousclassifications gradually become stable. The reason for this isthat both the training of RWN and the calculation of con-sistency loss depend on the labelled samples. The network isnot sufficiently trained when the number of labelled samples issmall, which leads to a decrease in accuracy for all categories.This indicates that there is room for further improvement ofthe algorithm.

\section{Conclusion}

In this paper, we proposed a novel ankle fracture classification algorithm to help CNN classification networks ground decisions on more sensible visual patterns. we used a deep learning algorithm to segment the tibia and fibula, and registered the segmentation mask containing the fracture with the normal bone mask to preprocess and classify the X-ray image of the ankle fracture. Besides, a new dataset was constructed based on the image data of fracture patients collected from Hospital. The pre-processing process based on bone segmentation and registration can effectively segment bone contours, and make full use of unlabeled data and adopt semi-supervised network training strategy to improve the accuracy of fracture classification. Finally, the feasibility of the proposed algorithm in the field of intelligent diagnosis of ankle fracture is verified by the experimental results. This work signifies a step forward towards using explainable artificial intelligence techniques to help humans understand and improve computer-aided diagnosis. The current method assumes a simple fracture presence and we plan to generalize it to deal with multiple, complicated, and compound fracture presences.

\bibliographystyle{IEEEtran}
\bibliography{main}

\end{document}